\documentstyle[preprint,aps]{revtex}
\begin{document}

\title{Convergent variational calculation of 
positronium-hydrogen-atom scattering lengths}

\author{Sadhan K Adhikari$^ \dagger$ and Puspajit Mandal$^{\dagger, \$}$}
\address{$^\dagger$Instituto de F\'{\i}sica Te\'orica, 
Universidade Estadual Paulista
01.405-900 S\~ao Paulo, S\~ao Paulo, Brazil\\
$^\$ $Department of Mathematics, Visva-Bharati, Santiniketan 731 235,
India}

\date{\today}
\maketitle

\begin{abstract}

We present a convergent variational basis-set calculational scheme for
elastic scattering of positronium atom by hydrogen atom in S wave.  Highly
correlated trial functions with appropriate symmetry are needed for
achieving convergence. We report convergent results for scattering lengths
in atomic units for both singlet ($=3.49\pm 0.20$) and triplet ($=2.46\pm
0.10$)  states.

{\bf PACS Number(s): 34.90.+q, 36.10.Dr}

\end{abstract}


\newpage

Lately, there has been interest in the experimental \cite{1} and
theoretical \cite{2,7,g,14,13,16,17} studies of ortho positronium (Ps)
atom
scattering by different neutral atomic and molecular targets. The Ps-H
system is theoretically the most simple and fundamental and a complete
understanding of this system is necessary before a venture to more complex
targets \cite{16,17,9}. There have been  R-matrix \cite{7},
close-coupling (CC) \cite{g,11a} and model-potential \cite{14,dh}
calculations
for Ps-H scattering. Here we present a convergent variational basis-set
calculational scheme for low-energy Ps-H scattering in S wave below the
lowest Ps-excitation threshold at 5.1 eV. Using this method we report
numerical results for scattering length of electronic singlet and triplet
states.

A recent study based on a regularized nonlocal electron-exchange model
potential \cite{13} yielded low-energy (total) cross sections in agreement
with experiment for Ps scattering by He \cite{13}, Ne \cite{16}, Ar
\cite{16} and H$_2$ \cite{17}.  For the Ps-H system the model-potential
results for S-wave singlet binding and resonance energies are in agreement
with accurate variational estimates \cite{18}. It would be interesting to
see if the model-potential result for the Ps-H singlet scattering length
 agrees with the
present convergent  calculation.

Because of the existence of two identical fermions (electrons) in the Ps-H
system, one needs to antisymmetrize the full wave function before
attempting a  solution of the scattering problem.
 The position vectors  of the electrons $-$ ${\bf
r_1}$ (Ps) and ${\bf r_2}$ (H) $-$ and positron (${\bf x}$) with respect
to (w.r.t.)
the massive proton at the origin are as shown in figure 1. We also use 
the position vectors   ${\bf s}_j= ({\bf
x+r}_j)/2$, $\rho_j = {\bf x-r}_j$, $j=1,2$, ${\bf r}_{12}={\bf
r}_1-{\bf r}_2$. 
The fully
antisymmetric state $\psi^{{\cal A}}_{\bf k}$ of Ps-H scattering is
given by $
|\psi^{\cal A}_{\bf k}\rangle={\cal A}_1|\psi^{1}_{\bf k}\rangle=(1\pm
P_{12})|\psi^{1}_{\bf k}\rangle = |\psi^{1}_{\bf k}\rangle \pm
|\psi^{2}_{\bf k}\rangle
$
where ${\bf k}$ is the incident momentum,  the antisymmetrizer ${\cal
A}_1$
is $(1+P_{12})$ for the 
singlet state and $(1-P_{12})$ for the triplet state with $P_{12}$ the
permutation operator of electrons 1 and 2. The function $\psi^1_{\bf k}$
refers to
the Ps-H
wave function with electron 1 forming the Ps as in figure  1 and
$\psi^2_{\bf k}$
refers to the same with the two electrons interchanged.

The full Ps-H Hamiltonian $H$ can be broken in two convenient forms as
follows
$H=H_1+V_1=H_2+V_2
$ where $H_1$ includes  the full kinetic energy and intracluster
interaction of H
and
Ps for the arrangement shown in figure 1 and $V_1$ is the sum of the
intercluster
interaction between H and Ps in the same configuration, $H_2$ and $V_2$
refer to the same quantities with the two electrons interchanged:
\begin{equation}
 V_1= \biggr[
\frac{1}{x}-\frac{1}{r_1}+\frac{1}{r_{12}}-\frac{1}{\rho_2} \biggr],\quad 
\quad
 V_2= P_{12}V_1=\biggr[
\frac{1}{x}-\frac{1}{r_2}+\frac{1}{r_{12}}-\frac{1}{\rho_1} \biggr] .
\end{equation}

The
fully antisymmetric state satisfies the Lippmann-Schwinger equation
\cite{ska}
\begin{equation}\label{2}
|\psi^{\cal A}_{\bf k}\rangle =|\phi^1_{\bf k}\rangle + G_1
V_1|\psi_{\bf k}^{\cal
A}\rangle.
\end{equation}
where the channel Green's function $G_1\equiv (E+i0-H_1)^{-1}$ and the
incident 
wave $|\phi^1_{\bf k}\rangle$ satisfies  $(E-H_1)|\phi^1_{\bf k}\rangle =
0. $ The incident Ps energy $E=6.8 k^ 2$ eV.
We are using atomic units (au) in which
$a_0=e=m=\hbar=1$,
where $e$ ($m$) is the electronic charge (mass) and $a_0$ the Bohr radius.
Using the definition  of the antisymmetrized state 
we rewrite (\ref{2})  as \cite{ska}
\begin{equation}\label{2x}
|\psi^{1}_{\bf k}\rangle =|\phi^1_{\bf k}\rangle + G_1 M_1
|\psi^{1}_{\bf k}\rangle
\end{equation}
\begin{equation}\label{3}
M_1 =V_1{\cal A}_1+(E-H_1)(1-{\cal
A}_1)\equiv  {\cal A}_1V_1+(1-{\cal
A}_1)(E-H_1).
\end{equation}
The properly symmetrized transition  matrix for elastic
scattering is defined by 
$\langle\phi^1_{\bf k} | T^{\cal A} | \phi^1_{\bf k} \rangle =
\langle\phi^1_{\bf k}|V_1|\psi^{\cal A}_{\bf k}\rangle = 
\langle\phi^1_{\bf k}|V_1 {\cal
A}_1|\psi^1_{\bf k}\rangle 
=
\langle\psi^1_{\bf k}|{\cal
A}_1V_1|\phi^1_{\bf k}\rangle $ \cite{ska}.
 A basis-set calculational scheme for the transition matrix can be
obtained from the following expression 
\cite{ska1}
\begin{eqnarray}\label{61}
\langle\phi^1_{\bf k} | T^{\cal A} | \phi^1_{\bf k} \rangle =
\langle\psi^1_{\bf k}|{\cal
A}_1V_1|\phi^1_{\bf k}\rangle+ \langle\phi^1_{\bf k}| {\cal
A}_1V_1|\psi^1_{\bf k}\rangle-
\langle\psi^1_{\bf k}|{\cal A}_ 1 V_1- M_1G_1{\cal A}_ 1 V_1 |\psi^1_{\bf
k}\rangle.
\end{eqnarray}
Using (\ref{2x}), it can be verified that 
(\ref{61}) is an identity if exact scattering wave fumctions $\psi^1_{\bf 
k}$ are used.
If approximate wave functions are used, (\ref{61}) 
is stationary w.r.t. small variations of $|\psi^1_{\bf
k}\rangle$ but not with  $\langle 
\psi^1_{\bf
k}|$. This one-sided variational property emerges because of the lack of
symmetry of the formulation in the presence of explicit antisymmetrization
operator ${\cal A}_1$. However, this variational property can be used to
formulate a basis-set calculational scheme with the following trial
functions \cite{ska1}
\begin{equation}\label{62}
 |\psi^1_{\bf
k}\rangle=\sum_{n=1}^N a_n|f_n\rangle, \quad \langle
\psi^1_{\bf
k}| = \sum_{m=1}^N b_m \langle f_m|.
\end{equation}
Substituting (\ref{62}) into (\ref{61}) and using this variational
property  w.r.t.  $| \psi^1_{\bf
k}\rangle$ we obtain \cite{ska1}
\begin{equation}\label{63}
\langle
\psi^1_{\bf
k}|=\sum_{m=1}^N  \langle \phi^1 _{\bf k}|
{\cal A}_ 1 V_1
|f_n
\rangle   D_{nm}\langle f_m|
\end{equation}
 \begin{equation}\label{5}
(D^{-1})_{mn}=
{\langle f_m  |{\cal A}_ 1 V_1- [{\cal A}_1V_1+(1-{\cal
A}_1)(E-H_1)]G_1{\cal A}_ 1 V_1 | f_n
\rangle}.
\end{equation}
Using the  variational form (\ref{63}) and definition $\langle\phi^1_{\bf
k} | T^{\cal A} | \phi^1_{\bf k} \rangle =
\langle\psi^1_{\bf k}|{\cal
A}_1V_1|\phi^1_{\bf k}\rangle$ we obtain the following basis-set
calculational scheme for the transition matrix
\begin{equation}\label{4}
\langle \phi^1 _{\bf k}|T^{\cal A}|\phi^1_{\bf k}\rangle  = \sum_{m,n=1} ^
N
{ \langle \phi^1 _{\bf k}|
{\cal A}_ 1 V_1
|f_n
\rangle   D_{nm}\langle f_m|{\cal A}_ 1 V_1| \phi^1_{\bf k}
\rangle }.
\end{equation}
(\ref{5}) and (\ref{4}) are also valid for the K matrix and in
partial-wave form where the momentum-space integration over the Green's
function $G_1$ should be performed with the principal-value prescription.

In the calculation, the basis  functions are taken in the following form
\begin{equation}\label{8}
f_m ({\bf r}_2,\rho_1,{\bf s}_1)=\varphi(r_2)\eta(\rho_1)e^{-\delta_m r_2
-\alpha_m\rho_1-\beta_m s_1 -\gamma_m(\rho_2+r_{12})-\mu_m
(x+r_1)} \frac{\sin(ks_1)}{ks_1}
\end{equation}
where 
$\varphi(r)=\exp(-r)/\sqrt \pi  $ and 
$\eta(\rho)=\exp(-0.5\rho)/\sqrt {8\pi} $
represent the H(1s) and Ps(1s) wave functions, respectively.
For elastic scattering the direct  Born amplitude  is zero and
the exchange correlation dominates scattering. To be consistent with this, 
the direct terms in 
the form factors  $\langle f_m |{\cal A}_1V_1
|\phi^1_{\bf k}\rangle$ and $\langle \phi^1 _{\bf k}|{\cal A}_1V_1 |f_n
\rangle$ are
zero with the above choice of correlations in the basis functions via
$\gamma_m$ and $\mu_m$. This property follows as the above function is
invariant w.r.t. the
interchange of  ${\bf x}$ and ${\bf r}_1$ whereas the remaining part of
the integrand in the direct terms  changes sign under this transformation.
A
proper choice of the correlation parameters $\gamma_m$ and $\mu_m$ is
crucial  for obtaining good convergence. 

In the following we specialize to the K-matrix formulation in S wave at
zero energy, when $\sin(ks_1)/(ks_1)=1$ in (\ref{8}). 
The useful matrix elements  of the present approach are explicitly written
as \cite{ska1}
\begin{equation}\label{7}
\langle \phi^1_p|{\cal A}_1V_1|f_n\rangle
= \pm \frac{1}{2\pi}\int 
\varphi({\bf r}_1)\eta({\rho}_2)\frac{\sin{ps_2}}{ps_2} 
[ V_1]
f_n({\bf r}_2,\rho_1,{\bf s}_1) 
d{\bf r}_2
d{\rho_1}
d{\bf s}_1
\end{equation}
\begin{equation}\label{52}
\langle f_m|{\cal A}_1V_1|\phi^1_p \rangle
= \pm \frac{1}{2\pi}\int 
f_m({\bf r}_1,\rho_2,{\bf s}_2) 
[V_1]
\varphi({\bf r}_2)\eta({\rho}_1)\frac{\sin{ps_1}}{ps_1} 
d{\bf r}_2
d{\rho_1}
d{\bf s}_1
\end{equation}
\begin{equation}\label{53}
\langle f_m|{\cal A}_1V_1|f_n\rangle
= \pm \frac{1}{4\pi}\int 
f_m ({\bf r}_1,\rho_2,{\bf s}_2) [ V_1]
f_n({\bf r}_2,\rho_1,{\bf s}_1)
d{\bf r}_2
d{\rho_1}
d{\bf s}_1
\end{equation}
\begin{eqnarray}\label{a}
\langle f_m | M_1 G_1  {\cal A}_1V_1| f_n \rangle \approx
-\frac{2}{\pi}
\int_0^\infty dp{\langle f_m|{\cal A}_1V_1|\phi^1_p \rangle \langle
\phi^1_p|{\cal A}_1V_1|f_n\rangle}
\end{eqnarray}
where the so called off-shell  term $ (1-{\cal A}_1)(E-H_1)  $
has been neglected for numerical simplification in this
calculation. This term is
expected to
contribute to refinement over the present calculation. 
In this convention  the on-shell K-matrix
element at zero energy is the scattering length: $a= \langle \phi_0^ 1|
K^ {\cal A}|\phi_0 ^ 1
\rangle.  $

All the matrix elements above can be evaluated by a method presented  in
\cite{19}. We describe it in the following for  $\langle
\phi_p^1 |{\cal A}_1V_1|f_n\rangle$ of (\ref{7}).  
By a transformation of variables from $({\bf r}_2, \rho_1,{\bf s}_1)$
to $({\bf s}_1,{\bf s}_2, {\bf x})$ with Jacobian $2^6$  and
separating the
radial and angular
integrations, the form factor (\ref{7})  is given by 
\begin{eqnarray}
\langle
\phi_p^1 |{\cal A}_1V_1|f_n\rangle
&=& \pm\frac{2^6}{16\pi^3}\int_0^\infty
s_2^2
ds_2  \frac{\sin(ps_2)}{ps_2}
\int_0^\infty s_1^2 ds_1e^{-\beta_ns_1}
\int _0^\infty x^2dx e^{-\mu _ n x} 
\nonumber \\ &\times& \int 
e^{-(ar_1+b\rho_1/2)}
e^{-(cr_2+d\rho_2/2)}
e^{-\gamma_n r_{12}}
[ V_1] d\hat s_1 d\hat s_2 d\hat x \label{13}
\end{eqnarray}
where $a=1+\mu_n$, $b=2\alpha_n+1$,  $c=1+\delta_n$ and 
$d=2\gamma_n+1$.
Recalling that ${\bf r}_j= 2{\bf s}_j -{\bf x}$, ${\bf r}_{12}=2({\bf
s}_1-{\bf s}_2)$, $\rho_j=2({\bf x}-{\bf s}_j), j=1,2$, 
 we employ  the following expansions of the exponentials
in (\ref{13}) 
\begin{eqnarray}\label{5x}
e^{-a |2{\bf s -x}|-b|{\bf x-s}|}=
\frac{4\pi}{sx}\sum_{lm}G_l^{(a,b)}(s,x)
Y^*_{lm}(\hat s)Y_{lm}(\hat x)
 \end{eqnarray}
\begin{eqnarray}\label{6x}
\frac{e^{-a|2{\bf s -x}|-b|{\bf x-s}|}}{|2{\bf s -x}|}=
\frac{4\pi}{sx}\sum_{lm}J_l^{(a,b)}(s,x)
Y^*_{lm}(\hat s)Y_{lm}(\hat x)
 \end{eqnarray}
\begin{eqnarray}\label{7x}
\frac{e^{-a|2{\bf s -x}|-b|{\bf x-s}|}}{|{\bf s -x}|}=
\frac{4\pi}{sx}\sum_{lm}K_l^{(a,b)}(s,x)
Y^*_{lm}(\hat s)Y_{lm}(\hat x)
 \end{eqnarray}
\begin{eqnarray}\label{8x}
\frac{e^{-a|{\bf s_1-s_2}|}}{|{\bf s_1 -s_2}|}=
\frac{4\pi}{s_1s_2}\sum_{lm}A_l^{(a)}(s_1,s_2)
Y^*_{lm}(\hat s_1)Y_{lm}(\hat s_2)
 \end{eqnarray}
\begin{eqnarray}\label{9x}
{e^{-a|{\bf s_1-s_2}|}}=
\frac{4\pi}{s_1s_2}\sum_{lm}B_l^{(a)}(s_1,s_2)
Y^*_{lm}(\hat s_1)Y_{lm}(\hat s_2)
 \end{eqnarray}
where the $Y_{lm}$'s are the usual spherical harmonics. 
Using  (\ref{5x}) $-$ (\ref{9x}) in 
(\ref{13})  we get 
\begin{eqnarray}\label{8z}
\langle
\phi_p^1 |{\cal A}_1V_1|f_n\rangle
&=& \pm{2^8}\int_0^\infty  e^{-\beta_n s_1}
ds_1\int_0^\infty ds_2 {\sin(ps_2)\over {p
s_2}}\int _0^\infty
dxe^{-\mu_nx}\sum_{l=0}^L(2l+1)
\nonumber \\
&\times&
\biggr[\frac{1}{x}G_l^{(a,b)}(s_1,x)
G_l^{(c,d)}(s_2,x)
B_l^{(2\gamma_n)}(s_1,s_2)
-
J_l^{(a,b)}(s_1,x)
\nonumber \\
&\times&
G_l^{(c,d)}(s_2,x)
B_l^{(2\gamma_n)}(s_1,s_2)
+
\frac{1}{2}G_l^{(a,b)}(s_1,x)
G_l^{(c,d)}(s_2,x)
\nonumber \\
&\times&
A_l^{(2\gamma_n)}(s_1,s_2)
- \frac{1}{2}G_l^{(a,b)}(s_1,x)
K_l^{(c,d)}(s_2,x)B_l^{(2\gamma_n)}(s_1,s_2)
\biggr].
\end{eqnarray}
where the $l$-sum is truncated to $l=L$.
This evaluation avoids  complicated angular integrations
involving  ${\bf s}_1$,
${\bf s}_2$ and   ${\bf x}$.
These integrals take a simple form requiring    straightforward  
numerical
computation of certain radial integrals only, which must, however,  be
carried out
carefully.
The functions $G_l$, $J_l$, $K_l$ etc. are easily calculated using  
(\ref{5x}) $-$ (\ref{9x}):
\begin{equation}G_l^{(a,b)}(s,x)=\frac{sx}{2}\int_{-1}^1dt P_l(t)
e^{-a |2{\bf s -x}|-b|{\bf x-s}|} \label{8xz}
\end{equation}
where $P_l(t)$ is the usual Legendre polynomial and $t$ is the cosine of
the angle between ${\bf s}$ and ${\bf x}$.
 The integrals (\ref{52}) and (\ref{53}) can
be evaluated similarly. 
For example 
\begin{eqnarray}\label{8zz}
\langle
f_m |{\cal A}_1V_1|\phi^1_p\rangle
&=& \pm{2^8}\int_0^\infty  e^{-\beta_m s_2}
ds_2\int_0^\infty ds_1 {\sin(ps_1)\over {p
s_1}}\int _0^\infty
dxe^{-\mu_mx}\sum_{l=0}^L(2l+1)
\nonumber \\
&\times&
\biggr[\frac{1}{x}G_l^{(c,d)}(s_1,x)
G_l^{(a,b)}(s_2,x)
B_l^{(2\gamma_m)}(s_1,s_2)
-
J_l^{(c,d)}(s_1,x)
\nonumber \\
&\times&
G_l^{(a,b)}(s_2,x)
B_l^{(2\gamma_m)}(s_1,s_2)
+
\frac{1}{2}G_l^{(c,d)}(s_1,x)
G_l^{(a,b)}(s_2,x)
\nonumber \\
&\times&
A_l^{(2\gamma_m)}(s_1,s_2)
- \frac{1}{2}G_l^{(c,d)}(s_1,x)
K_l^{(a,b)}(s_2,x)B_l^{(2\gamma_m)}(s_1,s_2)
\biggr]
\end{eqnarray}
\begin{eqnarray}\label{8zx}
\langle
f_m |{\cal A}_1V_1|f_n\rangle
&=& \pm{2^7}\int_0^\infty  e^{-\beta_n s_1}
ds_1\int_0^\infty ds_2 e^{-\beta_m s_2}\int _0^\infty
dxe^{-(\mu_n+\mu_m)x}\sum_{l=0}^L (2l+1)
\nonumber \\
&\times&
\biggr[\frac{1}{x}G_l^{(e,f)}(s_1,x)
G_l^{(g,h)}(s_2,x)
B_l^{(2\gamma_{mn})}(s_1,s_2)
-
J_l^{(e,f)}(s_1,x)
\nonumber \\
&\times&
G_l^{(g,h)}(s_2,x)
B_l^{(2\gamma_{mn})}(s_1,s_2)+
\frac{1}{2}G_l^{(e,f)}(s_1,x)
G_l^{(g,h)}(s_2,x)
\nonumber \\
& \times&  
A_l^{(2\gamma_{mn})}(s_1,s_2)
 -   \frac{1}{2}G_l^{(e,f)}(s_1,x)
K_l^{(g,h)}(s_2,x)B_l^{(2\gamma_{mn})}(s_1,s_2)
\biggr]
\end{eqnarray}
where $e=1+\delta_m+\mu_n$, $f=2\alpha_n+2\gamma_m+1$,
$g=1+\delta_n+\mu_m$,  $h=2\alpha_m+2\gamma_n+1$ and $\gamma_{mn}
=\gamma_m+\gamma_n$.

We tested the convergence of the integrals by varying the number of
integration points in the $x$, $s_1$ and $s_2$ integrals in
(\ref{8z}), 
(\ref{8zz}) and (\ref{8zx}) and 
the $t$ integral in (\ref{8xz}). The evaluation
of  (\ref{8z}),
(\ref{8zz}) and (\ref{8zx}) essentially involves four-dimensional
integration, which is performed with caution. 
The $x$ integration was relatively easy
and 20 Gauss-Legendre quadrature points appropriately distributed between
0 and 16 were enough for convergence. In the evaluation of integrals of
type (\ref{8xz})  40 Gauss-Legendre quadrature points were sufficient for
adequate convergence. 
The convergence in the
numerical integration over $s_1$ and $s_2$ was achieved with 300
Gauss-Legendre quadrature points between 0 and 12. The maximum value of
$l$ in the sum in (\ref{8z}), (\ref{8zz}) and (\ref{8zx}), 
$L$, is
taken to be 6  which is
sufficient for obtaining the convergence with the partial-wave
expansions (\ref{5x}) $-$ (\ref{9x}).  

\vskip .4cm {Table 1: Singlet ($a_s$) and triplet ($a_t$) Ps-H scattering
lengths 
for different  $L$ and $N$.} 
\vskip .2cm
\begin{centering}

\begin{tabular} {|c|   c c| c c| c c| c c |}
\hline 
 &  $L=$    &0 &$L=$ & 2 & $L=$  & 4& $ L=$
& 6    \\
$N$  & $a_t$ &$a_s$ & $a_t$ & $a_s$ &$a_t$   &$a_s$ &  $a_t$
& $a_s$  \\
  \hline
6&4.00 &3.61 &3.66 &4.00  &3.07 &3.75  &2.92  &3.74   
\\
7&3.98 & 4.11&3.25 &3.88   &2.99 &4.06  &2.85  & 4.00          \\
8 &3.93 & 4.12&3.35 &3.96 & 2.66&3.72  &2.54   &   3.73         \\
9 &3.97 &4.15 &3.42 &3.83&2.65& 3.92   &2.55  & 4.06          \\
10& 3.97& 4.22&3.42 & 3.92 &2.57 &  4.42  &  2.50 &         3.45 
\\
11&4.01 &3.82 &3.44 &3.91 &3.55&  3.75 & 2.67  &3.80            \\
12& & & & &2.54 &3.72 & 2.46 &3.73             \\
13& & & & &2.48&3.47& 2.46 & 3.49             \\
\hline
\end{tabular}

\end{centering}
\vskip 0.4cm

In a numerical calculation 
a judicial choice of the parameters in (\ref{8})
is needed for rapid convergence. As the variational method does not
provide a  bound on the result, the method could converge to a
wrong scattering length  if an  inappropriate (incomplete)
basis set is choosen. After some experimentation we find that to obtain
proper convergence   
the parameters
$\delta_n$ and
$\alpha_n$ should be taken to have both  positive and negative values,
$\gamma_n$ and $\mu_n$ should include values close to unity and  
$\beta_n$
should
have progressively increasing values till about 1.5. 
The results reported in this work are obtained with 
the following parameters for the 
functions $f_n, n=1,...,13$: $ \{\delta_n,\alpha_n,\beta_n, \gamma_n,\mu_n
\}\equiv
  \{-0.5,   -0.25,    0.3,    0.01,    0.02\},$ $
   \{-0.5,   -0.25,    0.5,    0.04 ,   0.02\},$ $ 
   \{-0.5,   -0.25,    0.7,    0.03 ,   0.06\},$ $ 
  \{ -0.2,   -0.1,    0.6,    0.2,    0.2\},$ $
   \{-0.1 ,   0.1 ,   0.8  ,  0.25,    0.25\},$ $ 
    \{0.2,   -0.2 ,   0.6 ,   0.35 ,   0.35\},
  \{ -0.1,   -0.1 ,   0.7 ,   0.4 ,   0.4\},
   \{ 0.15  ,  0.2  ,  0.8   , 0.5  ,  0.5\},$ $ 
    \{0.12 ,  -0.12   , 1  ,  0.7   , 0.7\},$ $ 
    \{0.2  ,  0.2  ,  1.2 ,   0.9  ,  0.9\},$ $ 
   \{ 0.1    ,0.2  ,  1.3   , 1  ,  0.7\},$ $ 
   \{ 0.2   , 0.1  ,  1.4  ,  0.7  ,  1\},$ $ 
   \{ 0.3  ,  0.15  ,  1.5 ,   1  ,  1\}$

In table 1 we show the convergence pattern of the present calculation
 w.r.t. the number of partial waves $L$ and
basis functions $N$ used in the calculation. The convergence is smooth
with increasing $L$. However, as the present calculation does not
produce a bound on the result, the convergence is not monotonic with
increasing $N$. 
The lack of an upper bound in this calculation is clearly revealed in
table 1 where the results do not decrease monotonically as the number of
terms in the trial wave function is increased.  The unbounded nature of
the results is consistent with the noted oscillation of the scattering
lengths
as $N$ increases. The oscillation is larger in the singlet
state where it is more difficult to obtain convergence.  Although, the
final result for  the largest $N$ and $L$ is supposed to be the most
accurate,
it is not known whether this result  is larger or smaller
than the exact one. Also, an estimate of error of this result is
not known. 
It is difficult to provide a quantitative measure of convergence. However,
from the noted fluctuation of the results for large $N$ for $L=6$ we
believe the error in the triplet scattering length to be less than 0.10
and in the singlet scattering length to be less than 0.20.
The final results of the
present
calculation are those for $N=13$ and $L=6$ with the above estimate of
error: $a_s= 3.49\pm 0.20$ au and
$a_t=2.46\pm
0.10$ au. 

The maximum number of functions ($N=13$) used in this
calculation is also pretty small, compared to those used in different
Kohn-type variational calculations for electron-hydrogen ($N = 56$) 
\cite{eh} and positron-hydrogen ($N\le 286$)\cite{ph} scattering.  Because
of the explicit appearance of the Green's function, the present basis-set
approach is similar to the Schwinger variational method.  Using the
Schwinger method, convergent results for electron-hydrogen \cite{ehs} and
positron-hydrogen \cite{phs} scattering have been obtained with a
relatively small basis set ($N\sim 10$). These suggest a more rapid
convergence in these problems with a Schwinger-type method.

Now we compare the present results with those of other calculations. 
While a  static-exchange  calculation by Hara and Fraser \cite{11a}
yielded
$a_s=7.28$ au and $a_t=2.48$ au 
a model potential calculation by Drachman and
Houston \cite{dh}
produced
$a_s=5.33$ au and $a_t=2.36$ au. 
A
22-state R-matrix calculation by Campbell et al. \cite{7} and a six-state
CC calculation by Sinha et al. \cite{g}
yielded $a_s= 5.20$ au, 
$a_t=2.45$ au and 
$a_s=5.90$ au, $a_t=2.32$ au, respectively.
These triplet scattering lengths  are in qualitative
agreement
with the
present
result: $a_t=2.46\pm 0.10$. However, the  singlet scattering lengths of
these calculations  
have not yet
converged. The disagreement
in the singlet channel shows  that it  is more difficult
to
get the converged result in the attractive singlet channel than in the
repulsive triplet channel. 
This is consistent with the common wisdom that a  scattering model at
low energies is
more sensitive to
the detail of an effective attractive interaction  
than to that of  a repulsive interaction.
The
nonconvergence of these results for the singlet scattering length was
conjectured before \cite{14}. 
Using a correlation between the S-wave scattering length and
binding
energy for the Ps-H system, the value $a_s=3.5$ au was predicted in that 
study \cite{14} in excellent agreement with the present
result: $a_s=3.49 \pm 0.20$.

To summarize, we have formulated a convergent basis-set calculational
scheme for S-wave Ps-H elastic scattering below the lowest inelastic
threshold using a variational expression for the transition matrix. We
illustrate the method numerically by calculating the singlet and triplet
scattering lengths: $a_s= 3.49\pm 0.20$ au and $a_t= 2.46\pm 0.10$ au.


The work is supported in part by the Conselho Nacional de Desenvolvimento -
Cient\'\i fico e Tecnol\'ogico,  Funda\c c\~ao de Amparo
\`a Pesquisa do Estado de S\~ao Paulo,  and Finan\-ciadora de Estu\-dos e
Projetos of Brazil.

\vskip 1 cm
{\bf Figure Caption:}

1. Different  position vectors for the Ps-H system w.r.t  the proton
(p) at the origin in arrangement 1 with electron 1 forming the Ps.

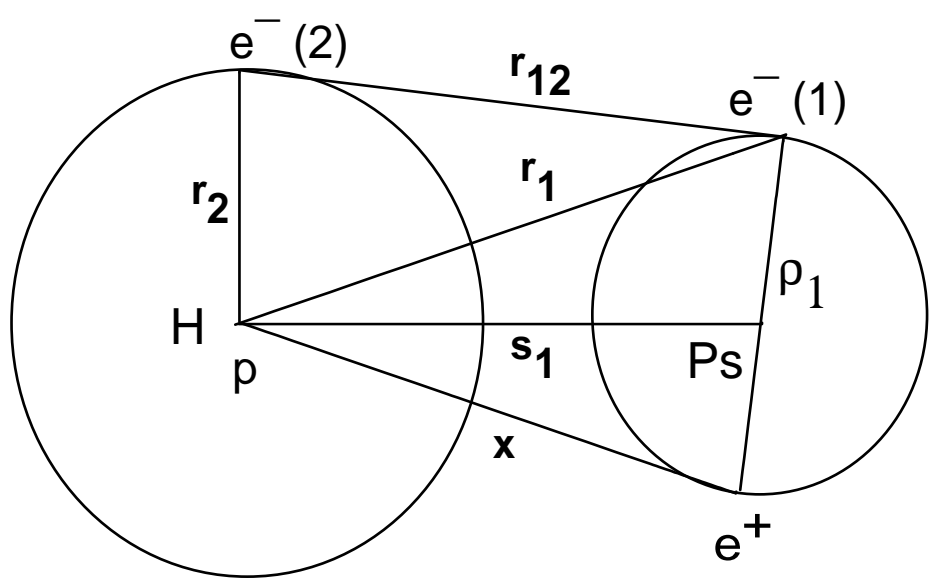
\end{document}